# Self-resonance effects for intrinsic Josephson junctions in $Nd_{2-x}Ce_xCuO_4$ films

T. B. Charikova, V. N. Neverov, M. R. Popov*, S. D. Popov, N. G. Shelushinina

M.N. Miheev Institute of Metal Physics of Ural Branch of Russian Academy of Sciences, 620041, Ekaterinburg, Russia

* e-mail: Popov_mr@imp.uran.ru

**Abstract**

To detect Josephson self-resonances, we used an original method, namely, studying the voltage $U_y$ at the Hall contacts in the $Nd_{2-x}Ce_xCuO_4/SrTiO_3$ film, where the $CuO_2$ planes are aligned along the longest side of the sample, perpendicular to the substrate. It is argued that the observed $U_y(j)$ oscillations are a set of Fiske steps in a layered superconductor system, indicating the manifestation of the ac-Josephson effect in a multilayer superconductor $Nd_{2-x}Ce_xCuO_4$ with a significant number of intrinsic Josephson junctions.

**Keywords**: Intrinsic Josephson junctions; $Nd_{2-x}Ce_xCuO_4$; Layered superconductors; Josephson vortices; Fiske steps; Swihart waves; Flux-flow regime.

## 1. Introduction

The first Josephson equation (*dc* Josephson effect) [1] for a superconductor–insulator–superconductor (SIS) tunnel junction establishes an interrelation between the dissipationless current flow (supercurrent) through the junction, $j$, and the phase difference, $\varphi$, of the superconductor wave functions: $j = j_c \sin\varphi$, where $j_c$ is the critical current. The periodic nature of this dependence leads to a possibility of oscillatory or resonant effects in real SIS systems under various external excitations (see monographs [2–5]).

It is well-known that the current-voltage characteristics (CVC) of Josephson tunnel junctions can display sharp current resonances. These features appear as a result of resonant interaction between *ac* Josephson current and electromagnetic waves that excited in the junction in the presence of magnetic field $B$ on the plane of the barrier and the voltage $V$ across the Josephson junction [2].

Due to the layered structure and the resulting strong anisotropy, high-temperature oxide superconductors can be viewed as a stack of Josephson tunnel junctions (intrinsic Josephson junctions) [6, 7], in which oscillations of various parameters of superconducting tunnel junctions are observed under the influence of a magnetic field.

We investigated the CVC at fixed magnetic fields on epitaxial films of electron-doped cuprate superconductor $Nd_{2-x}Ce_xCuO_4/SrTiO_3$, and in the flux-flow regime we observed the resonant jumps of tunneling current of intrinsic Josephson junctions.

We associate the maxima on CVC with internal resonances (self-resonances) of the Josephson current frequency with the frequency of Swihart waves - electromagnetic waves that arise at intrinsic Josephson junctions in parallel magnetic field (Fiske steps [8, 9, 10]).

## 2. Theoretical concepts

### *1. Self-resonance effects in magnetic field for single Josephson junction (after [11], [2])*

Let's consider a single tunnel junction between two superconductors (a contact between superconductors through a thin layer of insulator, SIS contact) in external magnetic field, *B,* parallel to the plane of the junction (see Fig. 1). Here the *x*-axis lies in the plane of the junction, and the magnetic field is directed along the *z* -axis. The effective magnetic thickness of the junction along the *y*-axis, where the current flows and the magnetic field exists, has the size $d = \boldsymbol{s} + \boldsymbol{2\lambda_L}$ with *s* being the thickness of the insulating layer of the tunnel junction, $\boldsymbol{\lambda_L}$ being the London penetration depth.

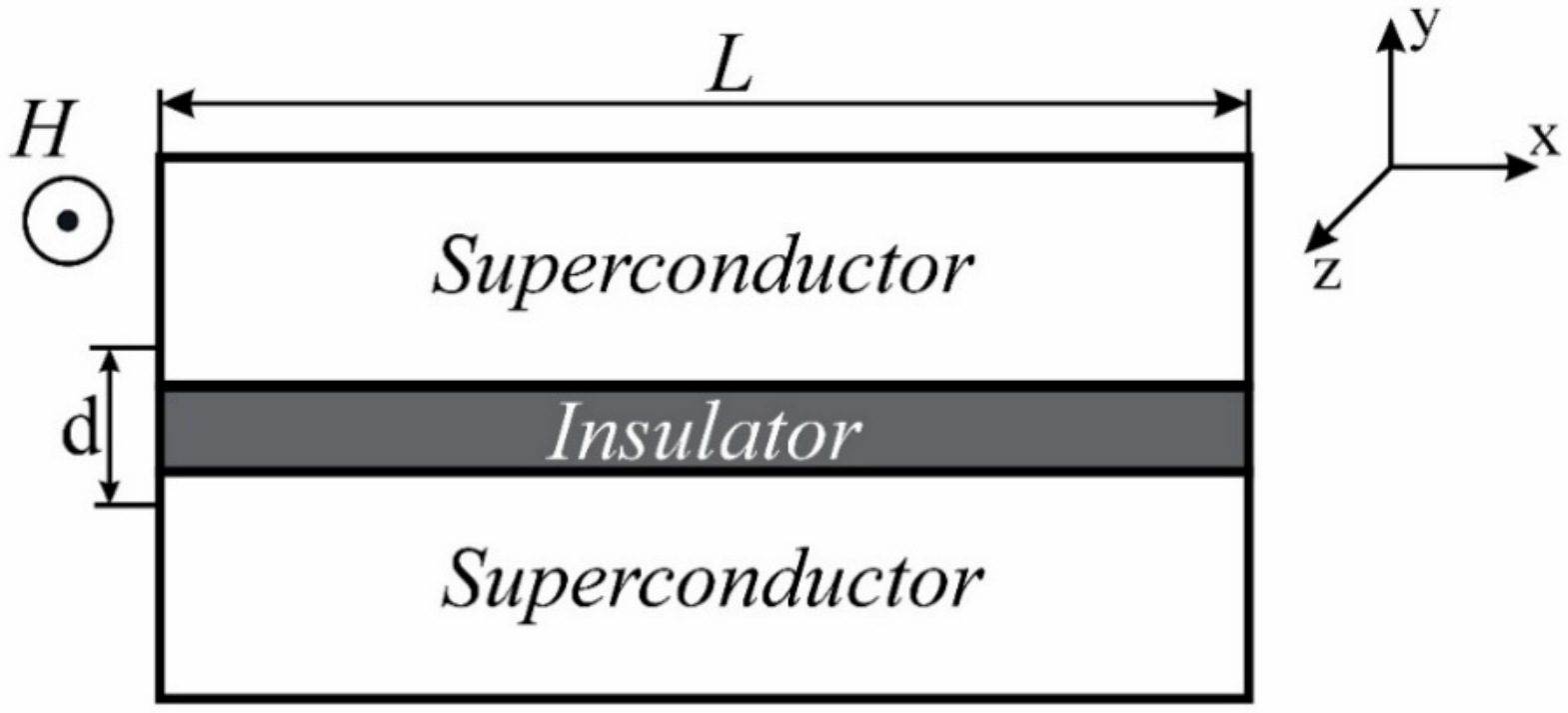

Figure 1: Josephson tunnel SIS junction of the length *L* placed in a magnetic field *H, d* - effective magnetic thickness of the junction.

Suppose that the Josephson junction is simultaneously under the influence of a magnetic field *B* applied parallel to the plane (*xz*) of the tunnel junction ($B||z$) and of a constant potential difference *V* across the junction. It was shown that in the absence of Josephson current, electromagnetic waves with velocity $c_0$ can spontaneously propagate along a contact between two superconductors separated by a dielectric layer. These are called Swihart waves [12] and

$$c_0 = c\left(\frac{s}{2\lambda_L\varepsilon}\right)^{1/2}, \qquad (1)$$

where $c$ is the velocity of light and $\varepsilon$ is the dielectric constant of the insulator.

Let us now consider the situation when, in the presence of voltage $V$ and magnetic field $B$ in the described configuration, there is a Josephson current. In this case, the phase difference has both spatial and temporal components and, due to the ac Josephson effect (second Josephson equation), a current inside the junction is oscillating at a Josephson frequency, $\omega_J$:

$$j = j_c\, sin[\varphi_0 + \omega_J t + kx]. \qquad (2)$$

Here $\omega_J = 2eV/\hbar$ and $k = 2eBd/\hbar$ is the wave vector of a traveling wave. The phase velocity, $\omega_J/k$, of this wave is equal to

$$v_0 = \frac{V}{Bd}. \qquad (3)$$

The complete solution of the problem in the presence of Josephson current, magnetic field $B$ and voltage $V$ leads to resonant features on the current-voltage characteristics of junction (Eck's maxima of the average current [13]). In a case of a junction of infinite length ($L = \infty$) only one resonance in the *I-V* characteristic is obtained.

At a given value of the external field, the maximum of this resonance occurs at a voltage $V_0$ such that $\omega_J/k = c_0$. This condition implies that the phase velocity of the Josephson current density distribution $v_0$ matches the phase velocity $c_0$ of the electromagnetic fields in the junction: $v_0 = c_0$, or, according to (3):

$$V_0 = c_0 Bd. \qquad (4)$$

The first experimental results exhibiting a behavior typical of a junction of infinite dimensions were reported by Eck, Scalapino, and Taylor [13]: well-defined peaks on *I-V* characteristic at different values of the applied magnetic field were observed for a long Pb-Pb junction (see Fig. 9.6 in [2]).

Until now, we have considered the contact to be infinite. In a finite contact of width $L$, standing electromagnetic waves can form if an integer number of half-waves fit within the distance $L$, i.e., $L = n\lambda/2$ or $k_n = n\pi/L$. This corresponds to the frequencies $\omega_n = \frac{c_0 n\pi}{L}$ and to the resonance potential values:

$$V_n = (\hbar c_0/2e)n\pi/L \text{ or } V_n = nc_0\Phi_0/2L, \qquad (5)$$

where $\Phi_0 = 2.07 \cdot 10^{-15} Wb$ is a magnetic flux quantum.

Hence, instead of one resonance for infinite contact, we obtain a set of discrete resonances. The calculation shows [14] that at the values of $V_n$ determined by condition (5), sharp maxima of the average current appear on the dependences $I(V)$.

In real experiments we obtain a step characteristic, the appearance of which is essentially determined by the magnetic field $B$, in the literature the term “Fiske steps” is accepted, named after the author who discovered such self-resonances on Sn-Sn and Pb-Sn tunneling junctions in 1964 [15], [16].

Experimental manifestations of Fiske steps in "short" Josephson junctions of various configurations have attracted considerable attention from researchers (see extensive references to early and recent works on this topic in [2] and [17], respectively).

As the contact width $L$ increases, the distance between adjacent Fiske peaks decreases according to formula (5). The peaks begin to overlap, and in the limit $L \to \infty$ only the maximum of Eck’s resonance occurs at a voltage $V_0$ corresponding to equation (4).

*2. Self-resonance effects in magnetic field for intrinsic Josephson junctions in multilayer superconductors (after [18], [19])*

The intrinsic Josephson effect, observed in high-*Tc* superconductors, is attributed to Josephson coupling between superconducting layers. Thus, the single crystals of a cuprate superconductor can be considered as the natural stacks of atomic scale intrinsic Josephson junctions [20]. Stacked Josephson junctions form multilayer transmission lines for electromagnetic (EM) waves. In theoretical models (see [19], [21] and references therein) the peculiarities of wave propagation and geometrical resonances in stacked Josephson junctions are phenomenologically described by models with the following parameters: $d$, the thickness of superconducting layers; $t$, the thickness of the tunnel barrier between the layers; $s = t + d$, the space periodicity of the stack; and $L$, the length of the stack.

The main difference between single and stacked junctions is the presence of multiple EM wave modes in the stack. Geometrical resonances in a stack correspond to formation of two-dimensional standing waves. The wave number along the *ab*-planes ($x$-axis) is $k_m = \pi m/L$, where $L$ is the length of the junctions and $m$ is the number of nodes in the standing wave. In the $c$-axis direction it is given by one of the eigen-modes, $k_n = \pi n/(N + 1)\, s$, $n = 1, 2, \dots , N$, where $N$ is the number of junctions in the stack. Each eigen-mode has a distinct propagation velocity which can be written as [18]:

$$c_n = c_0\left[1 - cos\left(\frac{\pi n}{N+1}\right)\right]^{-\frac{1}{2}}, n = 1,2,\ \dots\ , N, \quad (6)$$

where $c_0 = c\sqrt{ts/2\varepsilon\lambda_{ab}^2}$ is the Swihart velocity of a single junction in the stack, $\lambda_{ab}$ is the effective London penetration depth. This expression for $c_0$ is almost identical to that of Eq. (1) but considers the nonzero thickness of superconducting layers.

The lowest velocity $c_N$ characterizes the motion of a triangular fluxon lattice, which is energetically the most favorable fluxon mode at high magnetic fields. For *N*>>1, the lowest velocity may be written as [21]:

$$c_N \approx \frac{c}{2}\sqrt{\frac{ts}{\varepsilon\lambda_{ab}^2}} = \frac{c_0}{\sqrt{2}}. \quad (7)$$

In applied in-plane magnetic field Josephson vortices (fluxons) [22] enter into the junctions. In strong enough magnetic field fluxons form a regular fluxon lattice in a stack. Usually, a triangular lattice is most stable due to fluxon repulsion. Motion of fluxons leads to appearance of the flux-flow (FF) branch in current-voltage characteristics. Existence of EM waves in stacked junctions leads to excitation of geometrical resonances (Fiske steps) in the FF state [ 23].

Due to the *ac* Josephson effect, a current inside the junction is oscillating at a Josephson frequency, $\omega_J = (2e/\hbar)V$, where *V* is the *dc* voltage across the junction. Geometric resonances occur when the *ac*-Josephson frequency coincides with the frequency of one of the EM modes. The corresponding Fiske step voltage (per junction) for the resonant mode (*n, m*) is:

$$V_{mn} = \Phi_0 m c_n/2L, (m = 1,2,3, \dots ,\ \ n = 1,2, \dots , N). \quad (8)$$

As follows from (8) Fiske steps with a given *n* (almost) equidistant in voltage both for single [24], [25], [26] and stacked [27] junctions.

Thus, inside the resonant mode number *n* for a given value of $\Delta m$, the Fiske steps should be periodic with:

$$\Delta V_{FS} = N\bar{c}\Phi_0/2L, \quad (9)$$

for *N* junctions in the stack and $\bar{c} = \Delta m c_n$.

The strongest resonance occurs at the velocity matching (VM) condition, when the velocity of fluxons is equal to the velocity of electromagnetic waves what leads to appearance of the VM (Eck) step at the FF branch [26]. The VM voltage (per junction) is:

$$V_{VM} = c_n B s. \quad (10)$$

Irie and Oya [19] investigated the vortex dynamics in a stack of intrinsic Josephson junctions (IJJs) having length $L = 5\lambda_J$. The numerical simulations were performed by using coupled sine-Gordon model taking into account thermal fluctuations. In the absence of thermal fluctuations, the current–voltage characteristics show a series of Fiske steps corresponding to different cavity modes, excited by the collective vortex-flow depending on an external magnetic field.

However, in the presence of thermal fluctuations, it was found that only Fiske steps corresponding to the lowest-velocity cavity mode appear clearly and others become indistinct. This suggests that the out-of-phase mode becomes more stable than other modes due to fluctuations. The results are consistent with recent experimental observations in IJJs in the vortex-flow state.

Experimentally, clear Fiske steps have been observed in small intrinsic Josephson junction of mesoscopic $Bi_2Sr_2CaCu_2O_{8+\delta}$ single crystals under a magnetic field by several groups [18], [28], [29]. Surprisingly, the observed Fiske steps were due to the lowest mode velocity though many excitation modes are expected. These results may reflect the stability of the corresponding mode.

Katterwe et al. [21] studied Fiske steps in small $Bi_2Sr_2CaCu_2O_{8+x}$ mesa structures, containing only few stacked intrinsic Josephson junctions.The small number of junctions limits the number of resonant modes and allows precise identification of modes and velocities. Careful tuning of magnetic field allows one to observe a large variety of high-quality geometrical resonances, including superluminal ones with velocities exceeding the slowest velocity of electromagnetic waves.

**3. Sample and equipment**

Using the example of the manufactured epitaxial films $Nd_{2-x}Ce_xCuO_4/SrTiO_3$, we investigated the transverse voltage in the Hall configuration corresponding to the motion of the charge and vortex subsystems in orthogonal electric and magnetic fields. The magnetotransport properties of the layered electron-doped superconductor $Nd_{2-x}Ce_xCuO_4$ with $x$ = 0.145 and superconducting transition temperature $T_c$ = 15.7 K were studied. The sample is an epitaxial $Nd_{2-x}Ce_xCuO_4/SrTiO_3$ film grown in such a way that the *c*-axis of the NdCeCuO lattice is directed along the short side of the $SrTiO_3$ substrate (orientation $(1\bar{1}0)$) [30].

The film was fabricated in the form of a Hall bridge (Fig. 2), and the transverse voltage ($U$y) at the Hall contacts was measured at a constant current by reversing its sign and at the two orientations of the external magnetic field ($B^+$ and $B^-$). The electric current, ***j***, flowed along the $CuO_2$ planes (*x* direction), and the magnetic field vector, ***B***, was perpendicular to the substrate plane, directed along the $CuO_2$ planes, and orthogonal to the direction of the electric current $\boldsymbol{B} \perp \boldsymbol{j}$ (*z* direction). With this configuration of the external magnetic field and current in the flux - flow regime (see below), the vortices moved perpendicular to the current and magnetic field along the *c* axis (*y* direction). The coordinate system used here is shown in Fig. 2 (top left); for the geometry of the experiment, see also Appendix 1 (Fig. A1).

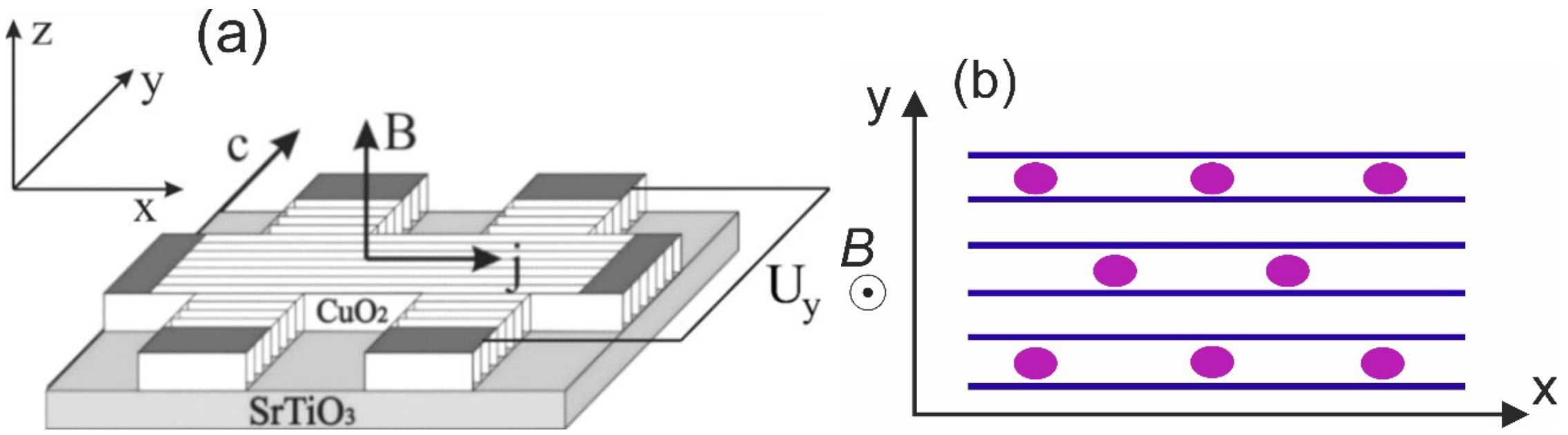

Figure 2: $Nd_{2-x}Ce_xCuO_4/SrTiO_3$ sample, $x = 0.145$, *c*-axis parallel to the short side of the film (a). Top view of the sample with a schematic representation of the centers of Josephson vortices (crimson ovals) located between the $CuO_2$ planes (b).

Table 1: Parameters of investigated $Nd_{2-x}Ce_xCuO_4$ film.

| $x$ | $L, \mu m$ | $w, \mu m$ | $s, nm$ | $\lambda_{c,}, nm$ | $\lambda_{ab}, nm$ | $\gamma$ | $d$, $nm$ | $\lambda_J, \mu m$ |
|---|---|---|---|---|---|---|---|---|
| 0.145 | 4300 | 800 | 0.6 | 450 | 25 | 18 | 50.6 | 100±5 |

Table 1 shows some parameters of the studied sample. Here *L* is the sample length (dimension in the *x*-direction, the width of the Josephson junction), *w* is the sample width (dimension in the *y*-direction) and *s* is the distance between the $CuO_2$ planes. Further, $\lambda_c$ and $\lambda_{ab}$ are the penetration depths for a magnetic field parallel to the *c*-axis and to the *ab*-planes, respectively, $\gamma = \lambda_c/\lambda_{ab} = \sqrt{\rho_c/\rho_{ab}}$ is the anisotropy parameter, $d = s + 2\lambda_{ab}$ is the Josephson vortex size in the *z* direction and $\lambda_J$ is the Josephson penetration depth.

The measurements were carried out on the original certified setup for measuring galvanomagnetic effects with a solenoid “Oxford Instruments” (Collaborative Access Center (CAC) "Testing Center of Nanotechnology and Advanced Materials") in magnetic fields up to 9 T at helium temperatures, $T = 1.7$ K and 4.2 K.

## 4. Experimental results and discussion

### *1. **C**urrent-voltage characteristics at T =* ***4.2 K***

Figure 3 shows dependences $U_y(j)$ in the studied sample for fixed magnetic fields in the range $B_{\parallel}^{-} = (0 - (-9))$T and for two directions of applied electric current, $j^+$and $j^-$, at $T = 4.2$ K. It is seen that in magnetic fields $|B| < 3$ T, with electric currents up to $j = 0.6$ mA, the voltage across the Hall contacts is not fixed. The finite voltage, corresponding to the transition to a resistive state caused by the motion of the Josephson vortex system (flux-flow resistive state), appears when the critical depinning current, $j_c(B_{\parallel})$, becomes less than the maximum applied external current (for more details, see our papers [31],[32]).

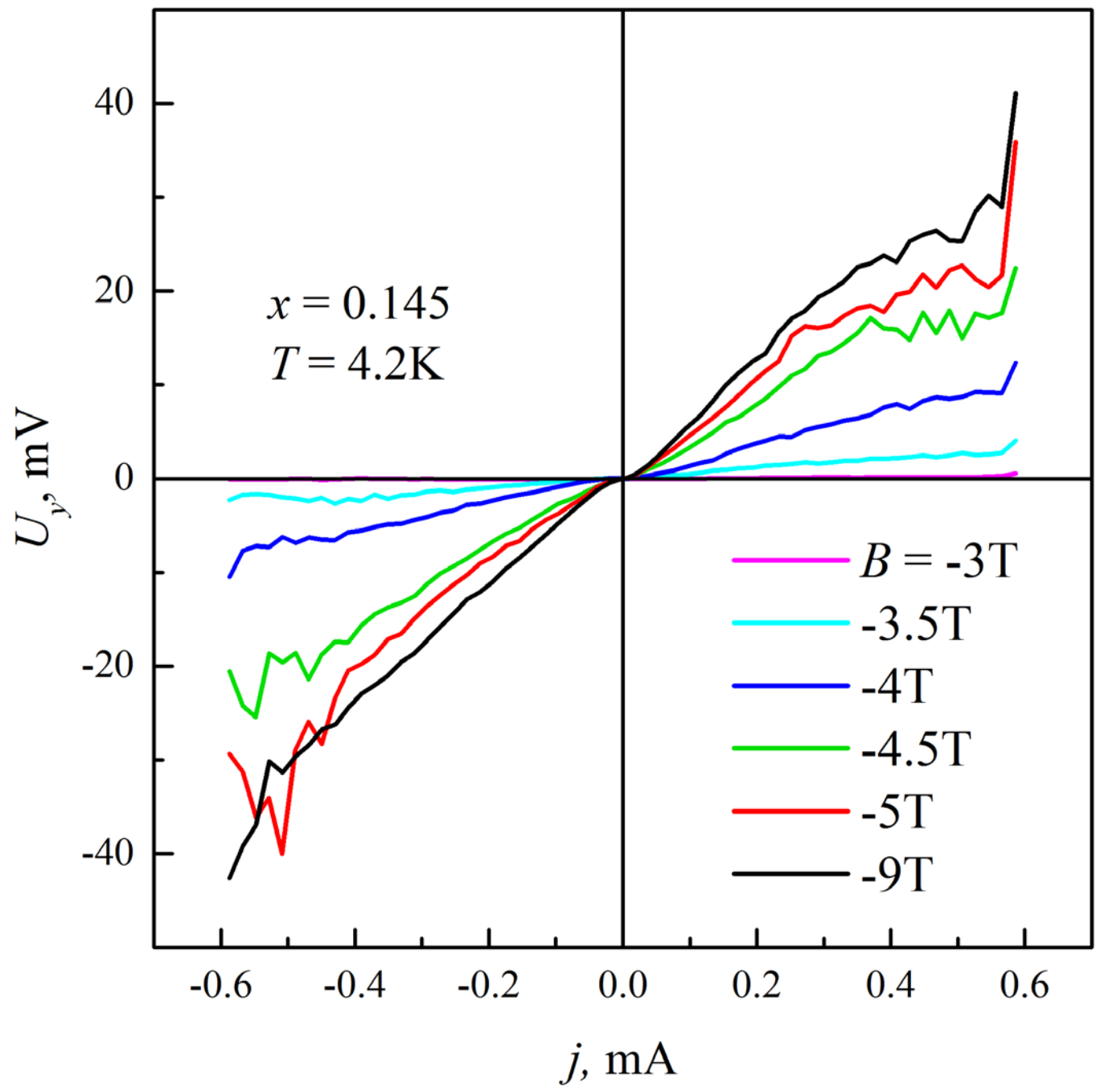

Figure 3: Dependences $U_y(j)$ for the studied sample in different magnetic fields ($B^-$) for two directions of applied current, $j^+$ and $j^-$, at $T = 4.2$ K.

In the current-voltage characteristics on Hall contacts in a constant magnetic field, we observe resonant current jumps, the amplitude of which increases with increasing field $B$ (see Fig. 3). Analysis revealed that the periodic structure is formed on the dependencies $U_y(j)$. We will discuss the physical significance of this fact later (see Section 4.2).

The presence of oscillatory structures on the curves $U_y(j)$ at $|B|$ = 3, 3.5 and 4 T is shown in Fig. 4 on an enlarged scale.

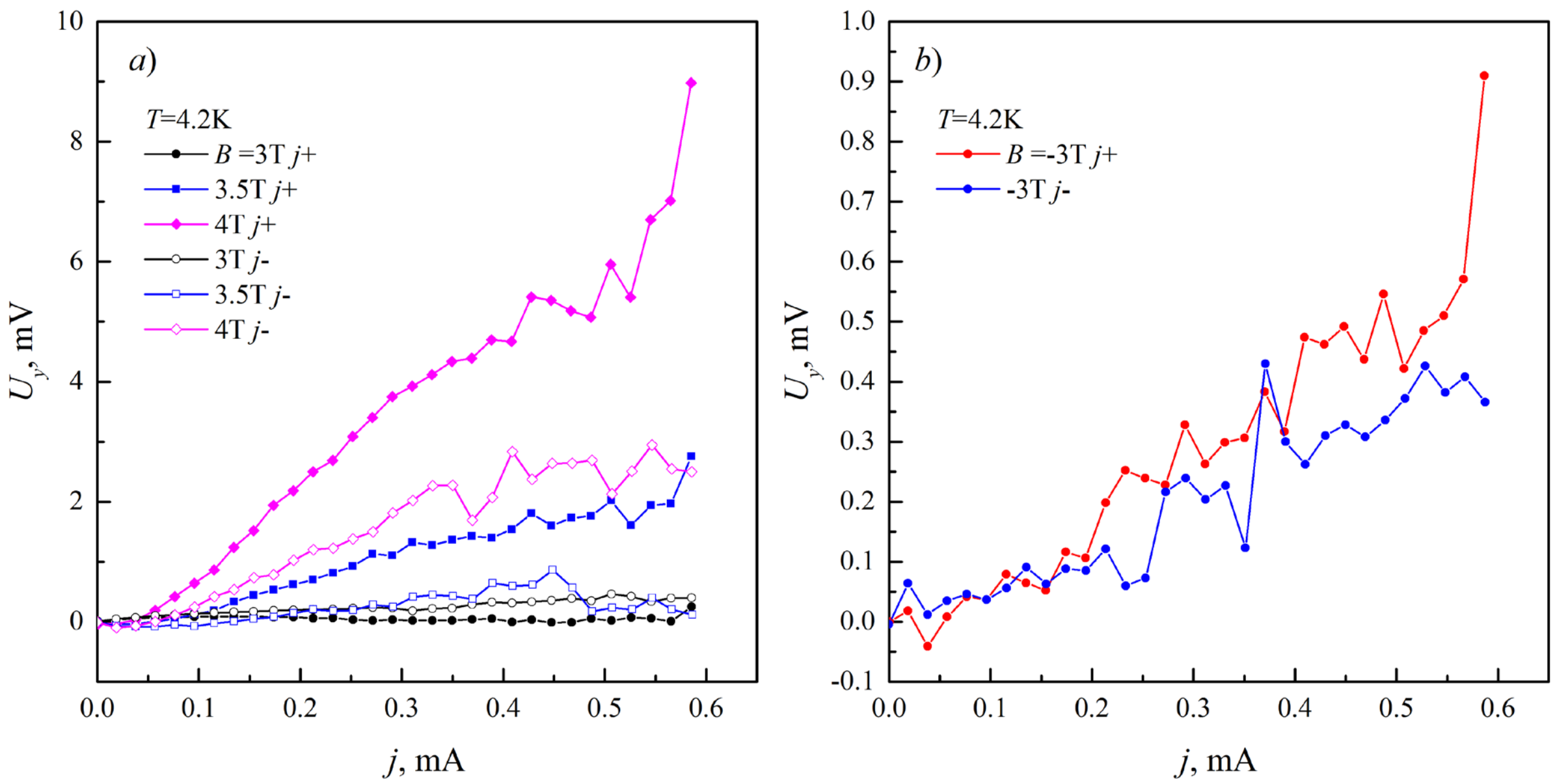

Figure 4: Dependences $U_y(j)$ of the sample studied at $T$ = 4.2 K (a) for $B$ = 3, 3.5 and 4 T and (b) for $B$ = 3 T on an enlarged scale to demonstrate resonant oscillations (see FFT results in Tables 2 and 3 as well as on Fig.5).

We performed a Fourier transform (FFT) analysis of $U_y(j^{\pm})$ dependences obtained at $T$ = 4.2 K for fixed magnetic fields in the range from 3T to 9T. An example of processing experimental data for $U_y(j^{\pm})$ to determine oscillation frequencies is presented in Appendix 2 for B = -3T.

The general results of the analysis: the set of frequencies (periods), $\Delta j$, found, are presented in Table 2. It is important that we observe a very limited set of periods $\Delta j^{\pm}$ (five values), and there is no systematic dependence of these values on the magnetic field: these five period values are randomly distributed in a set of different $B$ (see Table 3 and the graph of Fig. 5).

Table 2: Observed values of $\Delta j$ and their possible correspondence to resonant modes ($n$, $\Delta m$) and to Swihart velocities $c_n$ in equations (8) and (9).

| $\Delta j$, $10^{-5}$ A | 4.25 | 5.1 | 6.4 | 8.5 | 12.7 |
|---|---|---|---|---|---|
| Mode, ($n$, $\Delta m$) | (N,1) | SL | (0,1) | (N,2) | (0,2)/(N,3) |
| Velocity, $\bar{c}=mc_n$ | $1c_N$ | $c_{SL}$ | $1c_0$ | $2c_N$ | $2c_0/3c_N$ |

As shown in the Appendix 1 (see discussion in section 4.2), the periodicity in the applied current, $j$, is a manifestation of the periodicity (resonant effects such as Fiske steps) on the voltage, $U_H$, where

$$U_H = \frac{U_y(B) - U_y(-B)}{2} \tag{11}$$

is the antisymmetric (Hall) part of the total transverse voltage, $U_y$.

Since $U_H \sim j$, we have $\Delta j \sim \Delta U_H$, where the voltage period is determined by the formula (9):

$$\Delta U_H = N_{eff} \bar{c} \Phi_0 / 2L, \tag{12}$$

with $N_{eff}$ being the effective number of active junctions in the stack and $\bar{c} = \Delta m c_n$ for (*n*, *m*) resonant mode. Let us remind here that $c_0 = c\sqrt{ts/2\varepsilon\lambda_{ab}^2}$ is the Swihart velocity of a single junction in the stack and $c_N \approx c_0/\sqrt{2}$ is the lowest Swihart velocity for the stack with $N>>1$ [21]. For the parameters of the structure $Nd_{2-x}Ce_xCuO_4$ under study, $t = s = 6A^o, \lambda_{ab} = 250A^o, \varepsilon = 10$, we find $c_0 = 5.4 \times 10^{-3} c$; $c_N = 3.8 \times 10^{-3} c$ or $c_0 = 1.62 \times 10^6\ m/s$; $c_N = 1.14 \times 10^6\ m/s$.

For comparison, we present estimate of the Swihart velocity, $c_N = 0.83 \times 10^{-3} c$, for the structure $Bi_2Sr_2CaCu_2O_{8+x}$ ($s$=15.5 Å, $t$=12 Å, $\lambda_{ab}$=1700Å, $\varepsilon$ =25) from the work of Krasnov et al. [22]. Based on the expression (12), in Tables 2 and 3, we presented our assumptions about the identification of the observed periods, $\Delta j(\Delta U_H)$, according to the set of possible resonant modes (*n*, $\Delta m$).

Table 3: Distribution of oscillation modes obtained by FFT, according to the magnitude of the applied magnetic field.

| $B$, T | $\Delta j^+$ | $\Delta j^-$ |
|---|---|---|
| 3 | (0,2)/(N,3) | (0,2)/(N,3)<br>SL |
| 3,5 | (0,1) (N,3) | (0,2)/(N,3)<br>SL |
| 4 | (0,1) (N,3) | (0,2)/(N,3)<br>(0,1) |
| 4,5 | (N,2) | (0,2)/(N,3)<br>(0,1) |
| 5 | (0,2)/(N,3)<br>SL | (N,1) (N,2) |

| 9 | (0,1) | (N,2) SL |
|---|---|---|

In Fig.5. the same information is presented in the form of a graph, which shows the values of the periods $\Delta j^{\pm}$ for specific values of magnetic fields.

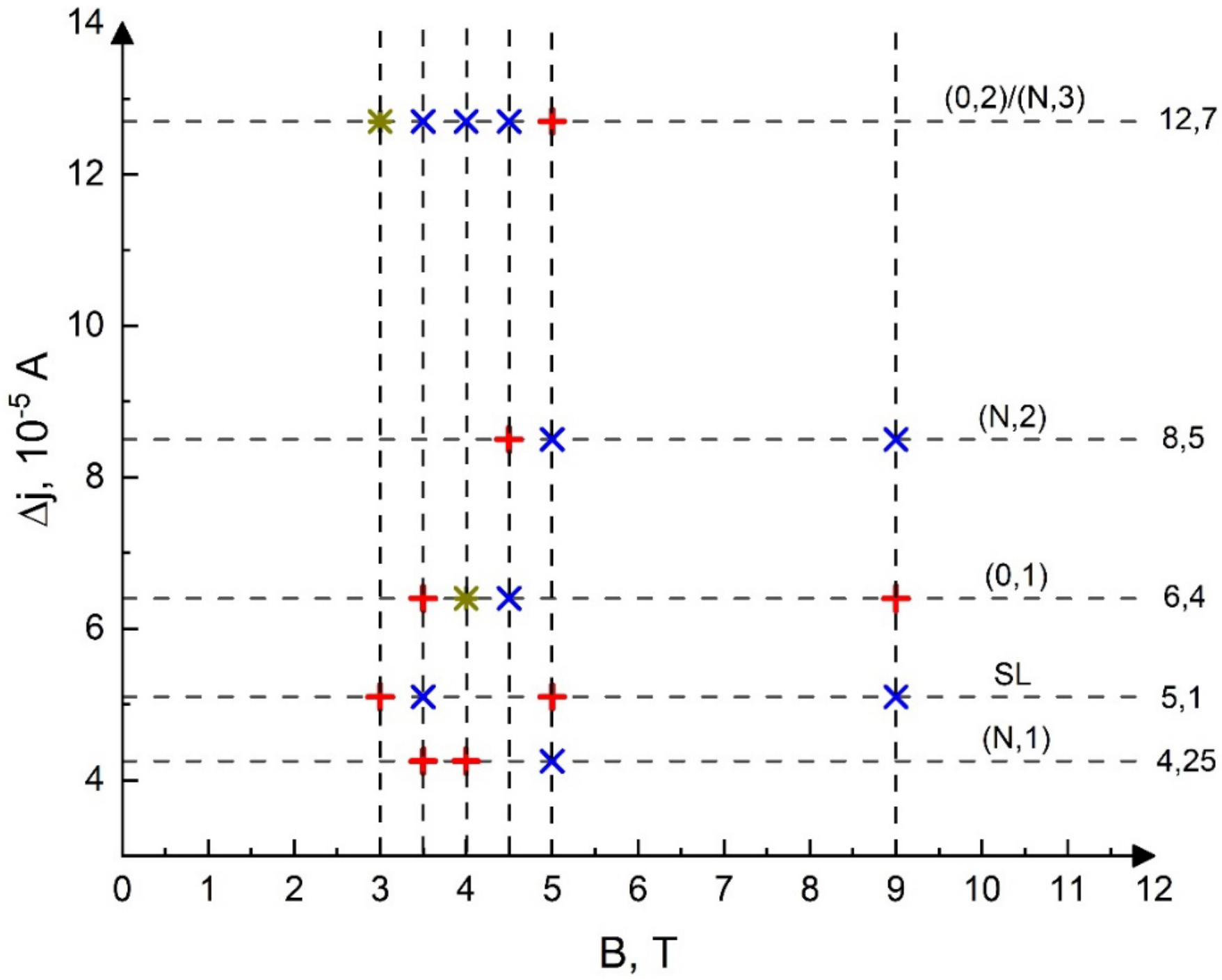

Figure 5: Diagram for the distribution of quantities $\Delta j^{\pm}$ on magnetic fields (+ - $\Delta j^{+}$;×- $\Delta j^{-}$; ✳ − $\Delta j^{+}$and $\Delta j^{-}$, simultaneously). The columns on the right show the corresponding modes (*n,* $\Delta m$) and values of $\Delta j$ in units of $10^{-5}$ A.

From Tables 2, 3 and Fig.5 we see that to describe the oscillatory behavior of the *I–V* characteristic of the studied sample it is almost sufficient to assume the manifestation of two fundamental modes (0,1) and (N,1) with Swihart velocities $c_0$ and $c_N$, as well as their second (0,2), (N,2) and third (N,3) harmonics. The only "extra" thing is the period $\Delta j = 5.1 \times 10^{-5}$A, which we on probation attributed to the superluminal (SL) mode in the terminology of Katterwe et al. [21] since the corresponding velocity $c_{SL} = 1.2\ c_N$.

It is possible, however, that this is an implementation of, for example, the (0,1) mode for a slightly different effective number of active layers, $N_{eff}$.

Table 4: Presents the values of the voltage periods recalculated using formula (A3): $\Delta U_H = R_{yx}\,\Delta j$, where, according to [32], the Hall resistance in the flux-flow regime, $R_{yx} = (15.0 \pm 2.5)\ \Omega$. From formula (12), with $c_0 = 1.62 \times 10^6\ m/s$, $c_N = 1.14 \times 10^6\ m/s$ and $L = 4{,}3 \cdot 10^{-3} m$ (see Table 1), we find an estimate for the effective number of active (phase-locked) intrinsic Josephson junctions in the studied sample: $N_{eff} \sim 10^4$.

Table 4: Periods of oscillations in the current-voltage characteristic converted to volts.

| Mode (n, Δm) | $\bar{c}=\Delta m c_n$ | $\Delta j$, $10^{-5}$ A | $\Delta U_H$, $10^{-4}$ V |
|---|---|---|---|
| (N,1) | $c_N$ | 4,25 | 6,4 |
| SL | $c_{SL}$ | 5,1 | 7,7 |
| (0,1) | $c_0$ | 6,4 | 9,6 |
| (N,2) | $2c_N$ | 8,5 | 13 |
| (0,2)/(N,3) | $2c_0/3c_N$ | 12,7 | 19 |

*2. Discussion of experimental results*

Let us consider in more detail the physical meaning of the oscillations observed in the current-voltage characteristics of the sample studied. Experimental measurements of the transverse voltage (voltage at the Hall contacts), $U_y$, as a function of the external current applied along the sample, $j$, were carried out (see Fig. 2).

In the studied film, $CuO_2$ layers are lined up in the direction of the long side of the sample, perpendicular to the plane of the substrate. Thus, the sample is a set of Josephson junctions (Jj's) as a stack of Jj's in the *c*-axis direction (*z*-axis in Fig. 2).

The applied current, $j$, together with the external magnetic field applied in *z* direction, creates a voltage $U_H \sim j \times B$ (as a part of the total voltage $U_y$). In turn, a combination of the external magnetic field, *B,* parallel to the planes of the junctions, and the voltage $U_H$, perpendicular to these planes, leads to the emergence of Swihart EM waves in the insulating layers of the tunnel Jj's between the $CuO_2$ planes.

Note also that

- in our system of long Jj's ($L \gg \lambda_j$), the magnetic field enters the sample in the form of Josephson vortices, the centers of which are located between the $CuO_2$ layers;

- the value $U_H$ becomes different from zero only in magnetic fields $B > B_c$ ($\approx$ 3T) after depinning of the Josephson vortices, in the flux-flow regime (see [32]).

Further, the voltage $U_H$, perpendicular to the Jj's plates, should lead to the *ac* Josephson effect if there is a component of the current $j_y$, directed perpendicular to the $CuO_2$ layers. The analysis carried out earlier [32], the results of which are presented in Appendix 1, shows that in our system a situation with a significant flow of current $j_y$ is realized due to some inhomogeneities in the sample and strong anisotropy of the structure.

Empirically this follows from a large value of the symmetrical with respect to *B* contribution, $U_R = R_{yy}j_y$, to the total transverse voltage $U_y$ (see Appendix 1). As calculations have shown, the contribution of $U_R(\sim j_y)$ in our samples amounts to 60% of the total voltage value.

In this case, it is important to us that the total voltage has a significant contribution proportional to the Josephson current $j_y$. Then the graph $U_y$vs $j$ in Fig. 3 can be considered as a graph $(a+bj_y)$ vs $U_H$, that is, as a current-voltage characteristic of the Josephson current, $j_y$, from the voltage $U_H$, applied perpendicular to the stack of Josephson junctions.

As a result, the observed periodic dependence $U_y(j)$ can be considered as a dependence $j_y(U_H)$, and we can assume that we see a self-resonance structure for the intrinsic Jj's corresponding to the Fiske steps.

Fiske steps in the flux-flow regime for the stacks of Josephson junctions have been experimentally studied in a number of works (see Table 5). As you can see, these are in the main research studies in small $Bi_2Sr_2CaCu_2O_{8+x}$ mesa structures, containing only few stacked intrinsic Josephson junctions. A small number of junctions limit the number of resonant modes and allow accurate identification of modes and velocities.

The general result of all the experimental works listed in Table 5 is the most stable manifestation of the out-of-phase mode with the lowest Swihart velocity**.** It follows that mode associated with the triangular lattice of Josephson vortices becomes more stable than other modes in a stack.

The emergence of such a stable collective system is due to the inductive coupling between separate junctions, which occurs if the superconducting layer thickness in a stack is equal or smaller than the London penetration depth. The mutual inductive coupling in stacks leads to phase-locking (voltage - locking) behavior with possible in-phase and out-of-phase oscillations in the adjacent junctions (see more about this topic, for example, in [27]).

As a result, both according to numerical calculations ([27], [19]) and experimentally (see Table 5.), in the structure of Fiske steps for the stacks of Jj's, mainly out-of-phase mode with the lowest Swihart velocity $c_N$ is explicitly manifested, though, in principle, a presence of many excitation modes was expected (see Eq. (6)).

The main feature of our film system $Nd_{2-x}Ce_xCuO_4$ is the naturally large number ($N\sim10^4$) of Josephson junctions involved in the stack. Nevertheless, we observe a very limited set of modes, which apparently indicates the action of a phase-locked pattern in our multilayer system. Thus, despite the significant number of junctions, it was possible to accurately identify resonant modes and velocities.

Table 5: The list of works where Fiske steps in the flux-flow regime for the stacks of Josephson junctions were observed ($N$ is the number of Jj's in the stack).

| Substance | $N$ | Refs |
|---|---|---|
| Nb/Al-AlO$_x$ /Nb | 7, 9 | [27] (1998) |
| $Bi_2Sr_2CaCu_2O_{8+x}$ (Bi-2212) | 5 | [18] (1999) |
| | ~60 | [33] (2001) |
| | ~120 | [29] (2005) |
| | ~300, ~600 | [30] (2006) |
| | 5 | [19]*) (2008) |
| Bi-2212 | 8, 11 | [21] (2010] |
| Bi (Pb)-2212 | ~15, ~56 | [34] (2011) |
| $Nd_{2-x}Ce_xCuO_4$ | ~$10^4$ | This work (2026) |

*) numerical simulations

## 5. Conclusions

An epitaxial film $Nd_{2-x}Ce_xCuO_4$, which is a stack of $N$ ($> 10^4$) intrinsic Josephson junctions, was investigated. Measurements of the current-voltage characteristics $U_y(j)$ ($U_y$ being voltage at the Hall contacts, $j$ being applied external current) in fixed magnetic fields $B$ (up to 9 T) were carried out.

In the flux-flow regime, the dependencies $U_y(j)$ show clearly defined oscillations with a set of periods that are independent on the magnetic field strength. We argue that the observed oscillations are a manifestation of Fiske steps in the dependence of the Josephson current on the voltage perpendicular to the plates of the Josephson junctions.

In agreement with theoretical calculations and experiments on mesoscopic structures (mainly on $Bi_2Sr_2CaCu_2O_{8+\delta}$), our system exhibits a very limited number of resonant modes (corresponding to Swihart velocities $c_0, c_N$ and $c_{SL} = 1.2c_N$), which indicates the formation of phase-locked (voltage-locked) state even for a stack with such a large number ($N_{eff} \approx 10^4$) of active Josephson junctions.

The observed resonant effects of Fiske step type in our system indicate the manifestation of the *ac* Josephson effect in a multilayer superconductor with a considerable number of intrinsic Josephson junctions.

The work was carried out within the framework of the state assignment of the Ministry of Science and Higher Education of the Russian Federation for the IMP UB RAS using the equipment of the Collaborative Access Center «Testing Center of Nanotechnology and Advanced Materials» of the IMP UB RAS.

## Appendix 1

### Geometry of Josephson junctions

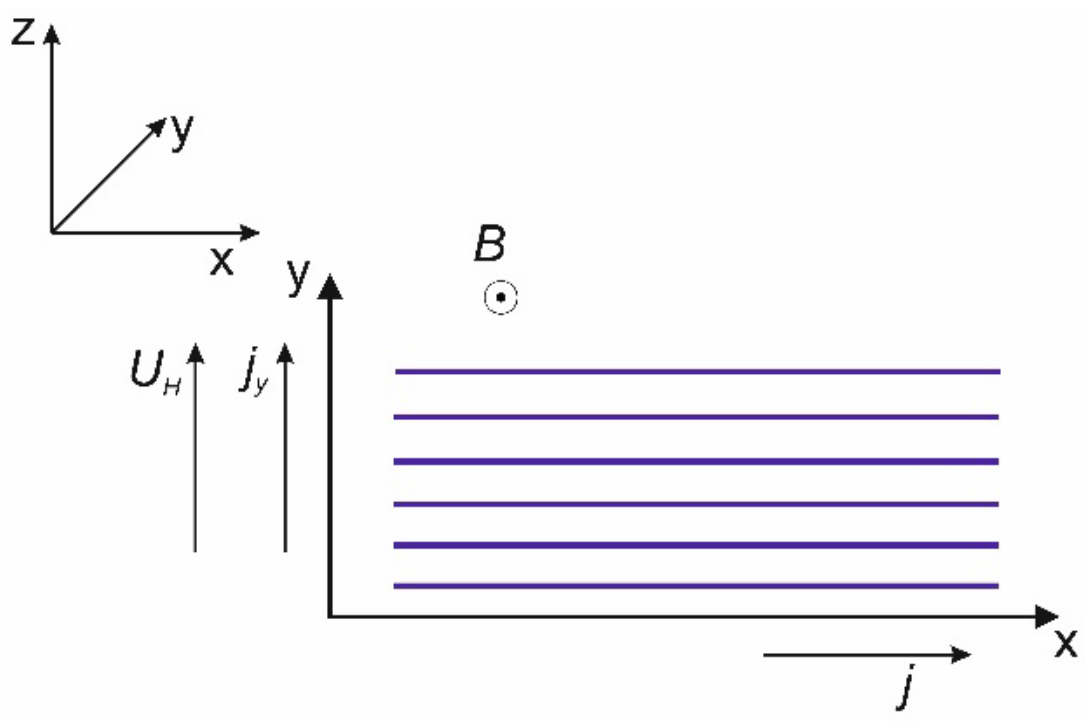

Figure A1: Top view of the measured sample.

We work in the geometry of Fig. 2. On Fig. A1 *j* is the applied current along the *x*-axis, the current $j_y = 0$ in the ideal case. The finite value of $j_y \neq 0$ may appear, for example, due to the asymmetry of the Hall contacts or, most likely, due to random inhomogeneities in the structure or violation of parallelism between layers and/or between the plane of the substrate and the deposited layers of the film. The X-ray studies of the films using a Dron-5M diffractometer ($K_{\alpha 1,\alpha 2}$-Co radiation) showed that in some cases there is a misorientation of the substrate plane and the deposited layers of no more than $0.6^0$. However, due to strong anisotropy in this compound ($\rho_c/\rho_{ab}$= 400), even such a deviation from parallelism can contribute to the conductivity between the layers.

In general case, for the total transverse voltage we have:

$$U_y\,(B) = R_{yx}\,(B) j_x + R_{yy}\,(B) j_y = U_H + U_R \tag{A1}$$

$$U_H = R_{yx}\,(B) j_x\,;\; U_R\; = R_{yy}\,(B) j_y\,,$$

where $U_H$ is the Hall contribution (odd in B), and $U_R$ is the contribution from the magnetoresistance (even in B).

Let's represent the variables in the form:

$$U_y\; = a + b j_y \rightarrow a = U_H; b = R_{yy}\,,$$

$$j \equiv j_x = \frac{U_H}{R_{yx}}.$$

Then we have

$$U_y \text{ vs } j \;\; \rightarrow \;\; \left(a + b j_y\right) \text{ vs } (U_H/R_{yx}\,). \tag{A2}$$

The oscillation period in volts can be obtained from the relation:

$$\Delta U_H = R_{yx}\,\Delta j. \qquad \text{(A3)}$$

Next, we have

$\frac{U_y\,(B)+U_y\,(-B)}{2J_x} = R_{yx}\,(B)$ – Hall resistance (usual result for determining the Hall coefficient);

$\frac{U_y\,(B)+U_y\,(-B)}{2\,j_y} = R_{yy}\,(B)$ - magnetoresistance.

For an isotropic crystal (a homogeneous medium) we have

$R_{yy} \sim R_{xx} \sim \rho_{xx}$ (resistivity), but $R_{yy} \neq R_{xx}$, because the geometric factors are different.

In our structure:

$R_{xx} \sim \rho_{xx} = \rho_{ab}$ for current in the plane $(ab)$;
$R_{yy} \sim \rho_{yy} = \rho_c$ for current along the *c*-axis.

Due to the strong anisotropy of the sample under study, $(\rho_c / \rho_{ab} = 400)$, it turned out that the *B*-symmetric contribution $(U_R)$ to the transverse voltage is about 60% of the total transverse voltage $U_y$ (see detailed analysis in our previous article [32]).

We argue that we see just the oscillations of the current $j_y$ (Josephson current), increased by the factor $R_{yy}(\sim\rho_c)$, in dependence on the perpendicular to intrinsic Jj's (to the $CuO_2$ planes) voltage $U_H$.

The use of the original (not averaged) experimental data in the main part of the article allows us to identify such subtle effects as the dependence of the critical current $j_c(B)$ (see [31]) and, most importantly, the resonant structures (Fiske steps) on the $j(U)$) dependences. Averaging over $(j, -j)$ and $(B, -B)$ "washes out" these effects (see [32]).

We believe that $U_y(j)$ CVC measurements are appropriate in our structures for the effects discussed above, since, due to strong anisotropy, the contribution to $U_y$ from magnetoresistance (symmetrical with respect to *B*) in our samples constitutes a significant part of the total voltage value. Furthermore, empirical studies have shown that this method is more sensitive than $U_x(j)$ CVC measurements.

## Appendix 2

### Direct and inverse Fourier transform of experimental data

In this appendix we describe a way for processing our experimental data to determine the oscillation frequencies of $U_y$ vs $j$, using the results for $B$ = - 3 T as an example. On Fig. 2a, the initial data (points) and their description using continuous curves (the process of "compacting" the data) presented, and Fig. 2b shows the selection of monotonic parts of the general dependencies $U_y(j^{\pm})$.

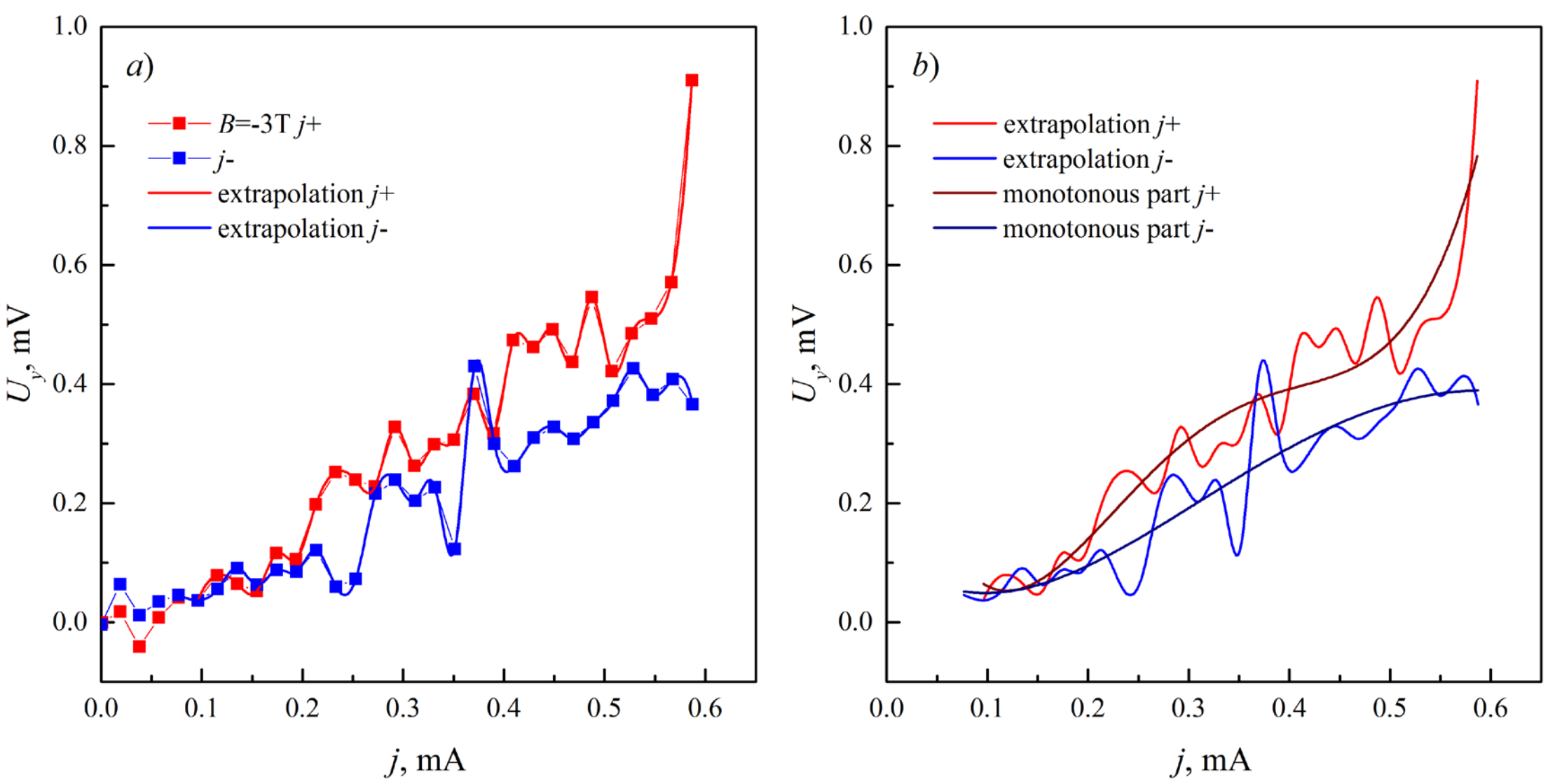

Figure A2: I-V characteristics of the studied sample for $B$ = - 3T at 4.2 K. (a) Description of the initial data by continuous curves and (b) highlighting the monotonic parts of the dependences $U_y(j^{\pm})$.

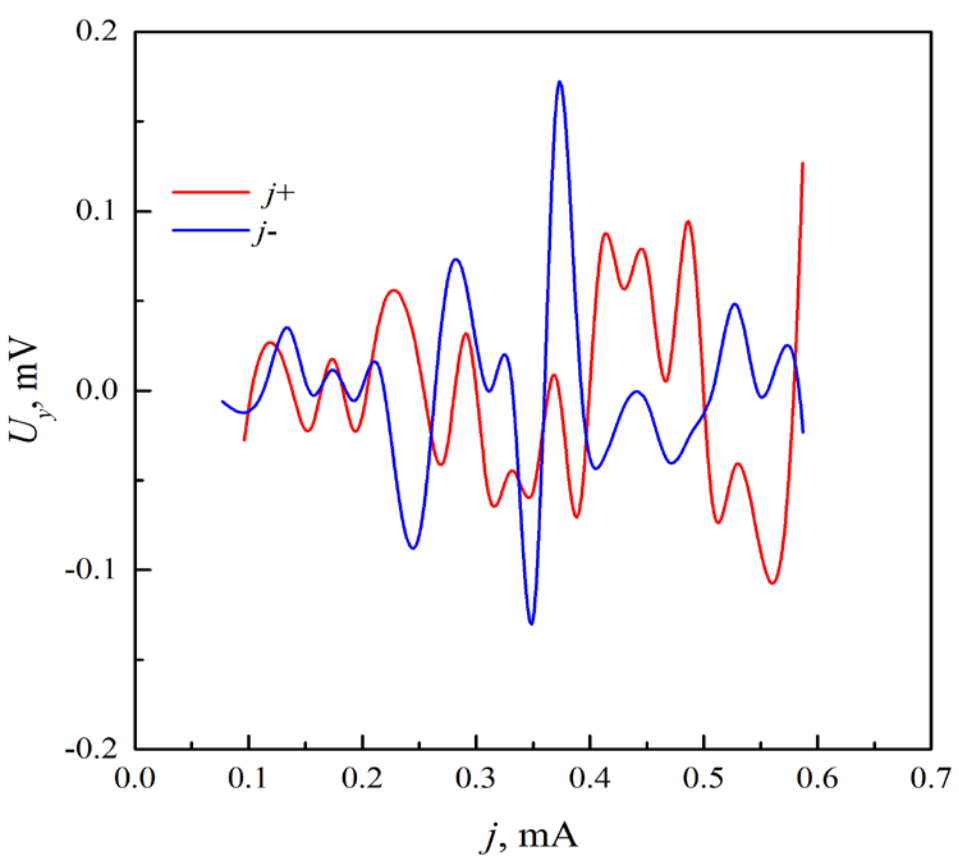

Figure A3: Dependences $U_y(j^{\pm})$ after subtracting monotonic components.

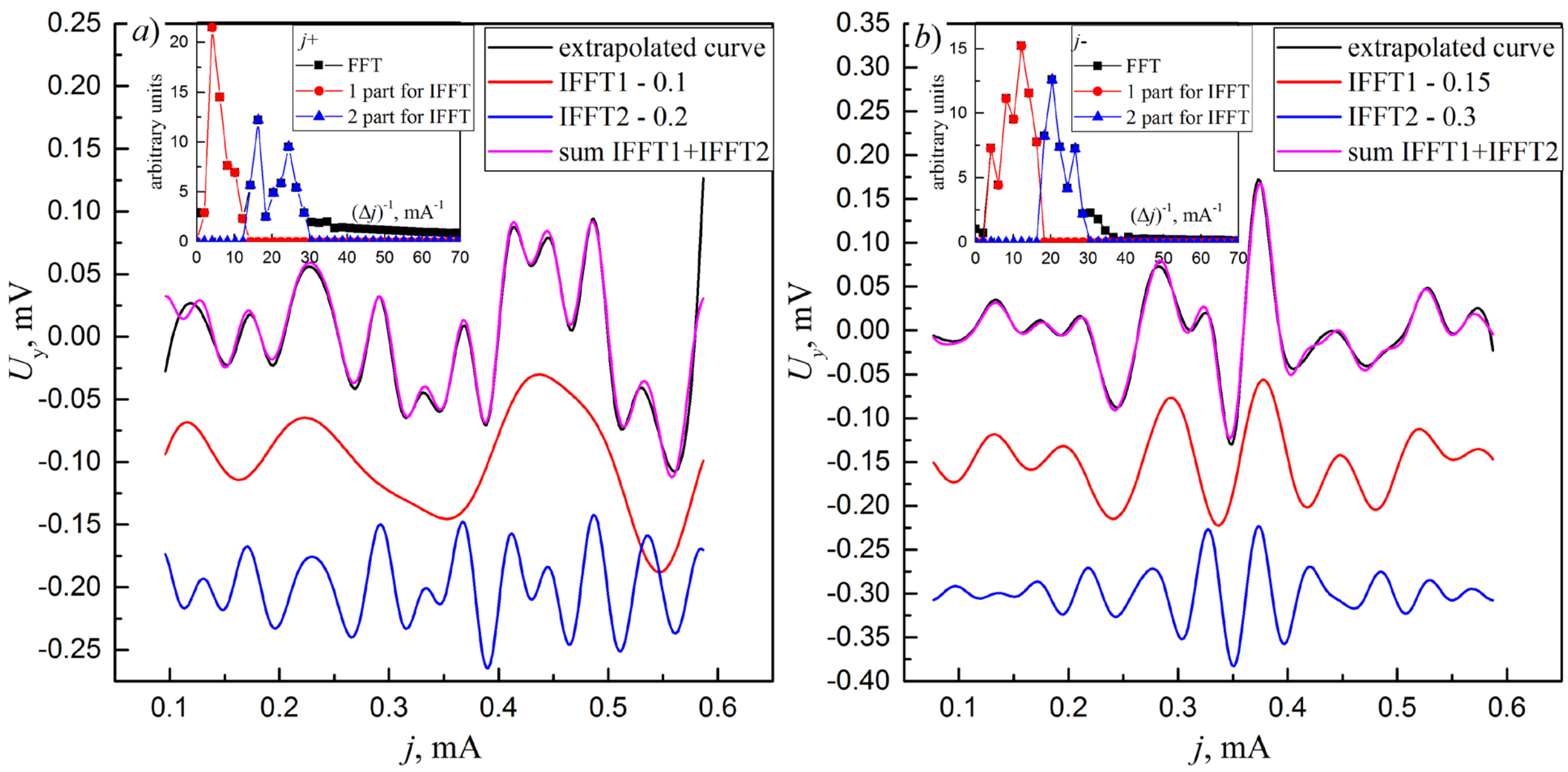

Figure A4: Results of direct (FFT) (inset) and inverse (IFFT) Fourier analysis of dependencies (a) $U_y(j^+)$ and (b) $U_y(j^-)$ obtained at $T$=4.2 K for fixed magnetic field $B$ = - 3T. For clarity, the IFFT1 and IFFT2 curves are shifted down on the x-axis.

The oscillation dependencies $U_y(j^\pm)$ obtained after subtracting the monotonic components are given in Fig. A3. In fact, the results of the direct (FFT) and, for verification, inverse (IFFT) Fourier transform of the data from Fig. A3 are presented in Fig. A4a, b, from which the self-consistency of the experimental data processing performed can be seen.